\documentclass[twocolumn,showpacs]{revtex4}
\usepackage{graphicx}
\usepackage{dcolumn}

\begin{document}


\title{Alpha decay and proton-neutron correlations}

\author{Kazunari Kaneko$^{1}$}
\author{Munetake Hasegawa$^{2}$}
\affiliation{
$^{1}$Department of Physics, Kyushu Sangyo University, Fukuoka 813-8503, Japan  \\
$^{2}$Laboratory of Physics, Fukuoka Dental College, Fukuoka 814-0193, Japan 
}

\date{\today}

\begin{abstract}

  We study the influence of proton-neutron ({\it p-n}) correlations on
 $\alpha$-decay width.  It is shown from the analysis of alpha $Q$ values 
 that the {\it p-n} correlations increase the penetration of the $\alpha$ particle
 through the Coulomb barrier in the treatment following Gamow's formalism,
 and enlarges the total $\alpha$-decay width significantly.
 In particular, the isoscalar {\it p-n} interactions play an essential role
 in enlarging the $\alpha$-decay width. 
 The so-called "alpha-condensate" in $Z \ge 84$ isotopes are related
 to the strong {\it p-n} correlations. 
\end{abstract}


\maketitle

  The $\alpha$ decay has long been known as a typical decay 
phenomenon in nuclear physics \cite{Gamov}.
Various microscopic approaches to estimating the formation amplitude
of the $\alpha$ cluster have been proposed \cite{Mang,Arima,Fliessback,Tonozuka}. 
The calculations \cite{Catara1,Pinkston,Catara2} showed that $J=0$ 
proton-proton ({\it p-p}) and neutron-neutron ({\it n-n}) 
pairing correlations cause substantial $\alpha$-cluster 
formation on nuclear surface. This suggests that the BCS approach 
with a pairing force offers a promising tool to describe the 
$\alpha$ decay. 
Proton-neutron ({\it p-n}) correlations are also 
significantly important for the $\alpha$-decay process 
in a nucleus \cite{Dodig1,Dodig2}. 
The effect of the {\it p-n} correlations on the $\alpha$-formation 
amplitude was studied by a generalization of the BCS approach
including the {\it p-n} interactions \cite{Delion1}, 
though it was shown that the enhancement of the formation 
amplitude due to the {\it p-n} interactions is small. 
The authors of Ref. \cite{Delion2} pointed out that continuum part 
of nuclear spectra plays an important role in formation of $\alpha$ cluster.
 On the other hand, a shell model approach including $\alpha$-cluster-model 
 terms \cite{Varga}
 gave a good agreement with the experimental decay width 
of the $\alpha$ particle from the nucleus $^{212}$Po. 
It is also interesting to investigate the effect of deformation
on the $\alpha$-decay width. 
{According to Ref.} \cite{Delion2}, the contribution of deformation
improves theoretical values for deformed nuclei such as $^{244}$Pu . 

 The {\it p-n} interactions are expected to become strong in 
$N\approx Z$ nuclei because valence protons and neutrons  
in the same orbits have large overlaps of wavefunctions \cite{Goodman}. 
In fact, this can be seen in peculiar behavior of the binding energy at $N=Z$.
 The double differences of binding energies are good indicators
to evaluate the {\it p-n} interactions \cite{Zhang,Brenner}. 
We have recently studied \cite{Kaneko1} various aspects of 
the {\it p-n} interactions in terms of the double differences of binding energies, 
using the extended $P+QQ$ force model \cite{Hasegawa}. 
The concrete evaluation confirmed that the {\it p-n} correlations 
become very strong in the $N\approx Z$ nuclei. 
It was shown in Ref. \cite{Kaneko2} that the 
isoscalar $(T=0)$ {\it p-n} pairing force persists over a wide range of
 $N>Z$ nuclei.
One of the double differences of binding energies was also discussed 
as a measure of $\alpha$-{\it particle} superfluidity in nuclei 
\cite{Gambhir,Hasegawa1}.
(We abbreviate the $\alpha$-like correlated four nucleons in a nucleus to
``$\alpha$-{\it particle}'' in italic letters.  The ``$\alpha$-{\it particle}''
is not a free $\alpha$ particle but a correlated unit in a nucleus.) 
We expect 
that the {\it p-n} correlations must play an important role in 
the barrier penetration of the $\alpha$ decay.

 Experimental evidence of the $\alpha$ clustering appears in the 
 systematics of alpha $Q$ values ($Q_{\alpha}$) \cite{Dussel},
 i.e., a large $Q_{\alpha}$ value coincides with 
 a large $\alpha$-decay width in the vicinity of the shell closures
 $Z=50$, $Z=82$ and $Z=126$ \cite{Roeckl}. 
The $Q_{\alpha}$ value is essentially important for penetration 
\cite{Geiger,Condon,Data,Delion3}. 
It is known that if experimental $Q_{\alpha}$ values are used 
 for the $\alpha$ decay between ground states,
 Gamow's treatment \cite{Gamov} describes qualitatively well 
 the penetration of the $\alpha$ particle through the Coulomb 
 barrier, even though the $\alpha$-{\it particle} is assumed to be 
 ``a particle'' in the nucleus. 
The penetration probability is expected to be sensitive 
 to the {\it p-n} component of the $Q_{\alpha}$ value.
 How much is the {\it p-n} correlation energy included in the $Q_{\alpha}$ value?
 What is the role of the {\it p-n} correlations in the barrier penetration? 
In this paper, we study these things and the effect of the 
{\it p-n} correlations on the $\alpha$ decay. 

  The total $\alpha$-decay width is given by 
the well-known formula \cite{Thomas}, 
\begin{eqnarray}
\Gamma= 2P_{L}\frac{\hbar^{2}}{2M_{\alpha}r_{c}}
g_{L}^{2}(r_{c}), \label{eq:1}
\end{eqnarray}
where $L$ and $M_{\alpha}$ denote, respectively, 
the angular momentum and the reduced mass of $\alpha$ particle, and $r_{c}$ 
channel radius. 
The $\alpha$ decay width depends on the two factors, 
the penetration factor $P_{L}$ and the $\alpha$ formation amplitude
 $g_{L}(r_{c})$.  
The $\alpha$ penetration is known as a typical phenomenon of 
``quantum tunneling'' in quantum mechanics. 

Since the $\alpha$-decay width depends sensitively 
upon the $Q_{\alpha}$ value, we 
first discuss the $Q_{\alpha}$ value, which is written in terms of the
 binding energy $B(Z,N)$ as follows:
\begin{eqnarray}
Q_{\alpha}(Z,N)=  B(Z-2,N-2) - B(Z,N) + B_{\alpha}, \label{eq:2}
\end{eqnarray}
where $B_{\alpha}$ is the binding energy of $^{4}$He. 
Experimental mass data show that the $Q_{\alpha}$ values
are positive for $\beta$-stable nuclides with mass number greater than about 150.
The $Q_{\alpha}$ value remarkably increases at nuclei
 above the closed shells, $N=$50, 82 and 126. 
This is attributed to dramatic increase of separation energy at the closed shells 
with large shell gap. 
The $Q_{\alpha}$ value becomes largest (about 10 MeV) above $N=128$,
and the $\alpha$ particle can penetrate a high Coulomb potential barrier
(which is 25 MeV for $^{212}$Po,
 and though it prohibits the emission of the $\alpha$ particle classically). 
 
\begin{figure}[b]
\includegraphics[width=8cm,height=7cm]{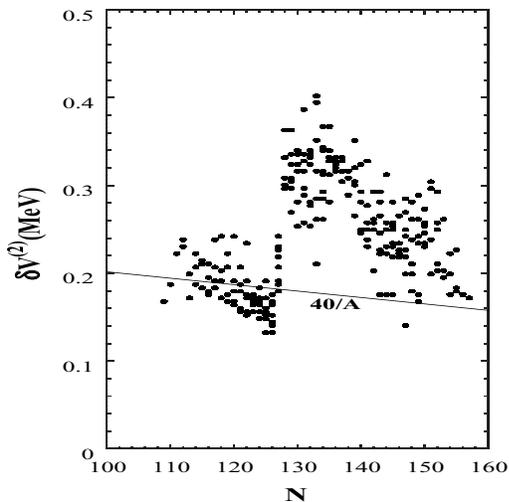}
  \caption{The experimental double difference of binding energies, $\delta V^{(2)}$
           for nuclei with proton number $Z=84\sim 100$ and neutron number 
           $N=110\sim 157$
           as a function of neutron number $N$
           along neutron chain.}
  \label{fig1}
\end{figure}

\begin{figure}[t]
\includegraphics[width=8cm,height=8cm]{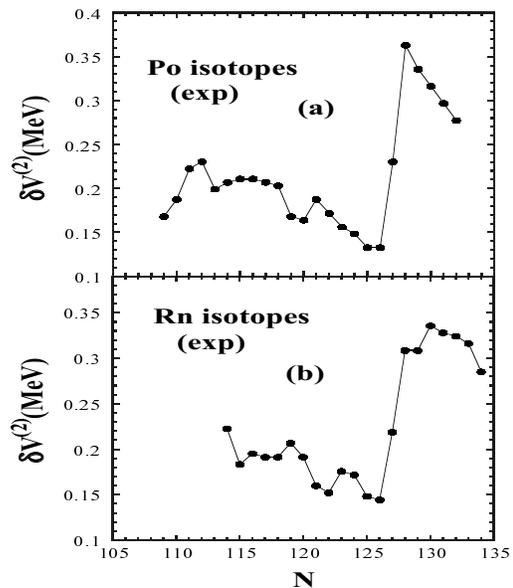}
  \caption{The experimental double difference of binding energies, 
              $\delta V^{2}$ for (a)the Po nuclei and (b)the Rn nuclei 
              as a function of neutron number $N$.}
  \label{fig2}
\end{figure}

 It has been shown in Refs. \cite{Zhang,Brenner,Kaneko1} that
the double difference of binding energies defined by
\begin{eqnarray}
\delta V^{(2)}(Z,N) &=&-\frac{1}{4}[B(Z,N)-B(Z,N-2)  \nonumber \\
   & - & B(Z-2,N)+B(Z-2,N-2)], \label{eq:3}
\end{eqnarray}
is a good measure for probing the {\it p-n} correlations. 
 We have recently studied global features of the {\it p-n} correlations
 in $A=40\sim 165$ nuclei by calculating the values of $\delta V^{(2)}$
 with the extended $P+QQ$ force model accompanied by an isoscalar ($T=0$)
 {\it p-n} force \cite{Kaneko2}. 
 The analysis has revealed that the $T=0$ {\it p-n} pairing interaction
 makes an essential contribution to the double difference of binding energies
 $\delta V^{(2)}$.
 The graph of $\delta V^{(2)}$ as a function of $A$ exhibits a smooth curve
 of $40/A$ on average and deviations from the average curve $40/A$ are small,
 in the region of the mass number $80<A<160$
  \cite{Brenner,Kaneko1,Kaneko2}. 
 Observed $\delta V^{(2)}$ can be reproduced with the use of semi-empirical
 mass formula based on the liquid-drop model and
 the symmetry energy term is the main origin of $\delta V^{(2)}$ \cite{Leander}.
 However, the parameters of the liquid-drop model are adjusted to experimental 
 binding energies and the liquid-drop model does not give sufficient 
 information about correlations in many-nucleon systems \cite{Gambhir1}. 
 Our analysis \cite{Kaneko1,Kaneko2} indicates that the symmetry energy
 in the liquid-drop model is attributed dominantly to the $J$-independent $T=0$ 
 {\it p-n} pairing force when considering in the context of correlations. 
 (In SO(5) symmetry model, the contributions of the $J$-independent $T=0$ 
 {\it p-n} pairing and the $J=0$ isovector ($T=1$) pairing forces to $\delta V^{(2)}$ 
 are estimated to be 73\% and 27\%, respectively. )

 In Fig. \ref{fig1}, we show the values of $\delta V^{(2)}$ observed
 in isotopes with proton number $Z=84\sim 100$. 
 This figure displays dramatic deviations from the average
curve $40/A$ in contrast with that for $80<A<160$ shown in Ref. \cite{Kaneko2}.
 The large deviations, however, seem to be different from those 
 in $N=Z$ nuclei with $N<30$ discussed in Refs. \cite{Brenner,Kaneko1},
 because nuclei with $N>128$ which has a large number of excess neutrons
 are in a very different situation from the $N=Z$ nuclei. 
The large deviations from the average curve $40/A$
for $N>128$ cannot be explained by only 
the symmetry energy or the $J$-independent $T=0$ {\it p-n} pairing force
which smoothly varies with nucleon number and is almost insensitive 
to the shell effects. 
 Nuclei with large $\delta V^{(2)}$ in Fig. \ref{fig1} are simply 
 those with short half-lives (i.e., large $\alpha$-decay widths),
  above the double-closed-shell nucleus $^{208}$Pb.
The plots of $\delta V^{(2)}$ for the Po and Rn nuclei extracted
from Fig. \ref{fig1} are shown in Figs. \ref{fig2}(a) and \ref{fig2}(b).
 We can see dramatic changes of $\delta V^{(2)}$ at $N=128$
 both in the Po and Rn nuclei. 

\begin{figure}[t]
\includegraphics[width=8cm,height=7cm]{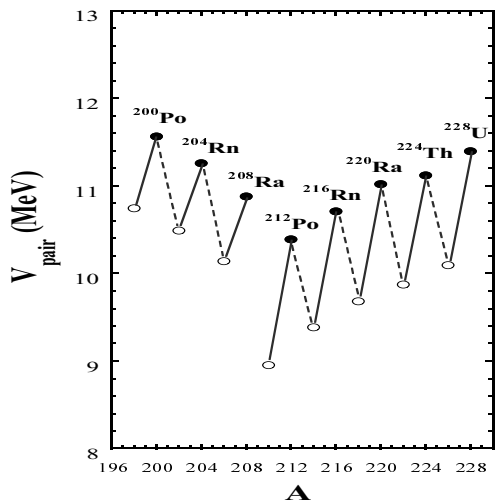}
  \caption{Even-odd staggering along the alpha-line as a function of mass number. 
  The solid circles denote $V_{pair}^{even}(A)$ for even pair number and the open circles 
  $V_{pair}^{odd}(A-2)$ for odd pair number.}
  \label{fig3}
\end{figure}

It is important to note that $\delta V^{(2)}$ is largest for $^{212}$Po 
with one $\alpha$-{\it particle} in Fig. \ref{fig2}(a) and $^{216}$Rn 
with two $\alpha$-{\it particle}s in Fig. \ref{fig2}(b) (and so on),
outside the double-closed-shell core $^{208}$Pb. 
The peaks are intimately related to 
 the even-odd staggering of proton or neutron pairs 
 from the ``alpha-condensate" point of view
discussed by Gambhir {\it et al.}\cite{Gambhir}.
They defined the following quantities for the correlations between pairs:
\begin{eqnarray}
V_{pair}^{even}(A) & = & \frac{1}{2}(B(Z-2,N)+B(Z,N-2)) \nonumber \\
& & - B(Z,N), \label{eq:6} \\
V_{pair}^{odd}(A-2) & = & B(Z-2,N-2) - \frac{1}{2}(B(Z-2,N) \nonumber \\
& & + B(Z,N-2)). \label{eq:7}
\end{eqnarray}
In Fig. \ref{fig3}, $V_{pair}^{even}(A)$ for even pair number and 
$V_{pair}^{odd}(A-2)$ for odd pair number are plotted along the
alpha-line nuclei which can be regarded as many {\it $\alpha$-particles}
outside the core $^{132}$Sn or $^{208}$Pb. 
The magnitude of the staggering corresponds to
the double difference of binding energies, {\it i.e.}, 
$\delta V^{(2)}(Z,N)=(V_{pair}^{even}(A)-V_{pair}^{odd}(A-2))/4$. 
We can see that
 the magnitudes (about 1.33 MeV) for the isotopes with $N>126$ are almost twice
as large as those (about 0.78 MeV) for the lighter isotopes with $N<126$.
Thus the peaks of $\delta V^{(2)}$ observed in Fig. \ref{fig3} are related to
the superfulid condensate of $\alpha$-{\it particle}s proposed by Gambhir {\it et al.}
The strong $\alpha$-like $2p-2n$ correlations, which enlarge $\delta V^{(2)}$, are 
important for $A>208$ nuclei with large $\alpha$ decay widths. 

\begin{figure}[h]
\includegraphics[width=5.0cm,height=4.0cm]{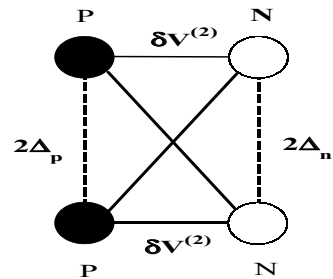}
  \caption{Schematic illustration of the {\it p-p} ({\it n-n}) correlations 
           and the {\it p-n} correlations in a correlated unit, 
           $\alpha$-{\it particle}.}
  \label{fig4}
\end{figure}

As mentioned earlier, the barrier penetration in the $\alpha$ decay
is very sensitive to the $Q_{\alpha}$ value. 
It is interesting to examine how much the {\it p-n} correlation energy is 
included in the $Q_{\alpha}$ value. What roles do the {\it p-n}
interactions play in the $\alpha$ decay? 
When we use the one proton (neutron) separation energy 
$S_{p}(S_{n})$, 
\begin{eqnarray}
S_{p}(Z,N)= B(Z,N) - B(Z-1,N), \label{eq:4} \\
S_{n}(Z,N)= B(Z,N) - B(Z,N-1), \label{eq:5}
\end{eqnarray}
and the three point odd-even mass difference for proton and neutron 
\begin{eqnarray}
\Delta_{p}(Z,N) & = & \frac{(-1)^{Z}}{2}( B(Z+1,N)- \nonumber \\
   & {} & 2B(Z,N)+B(Z-1,N)), \label{eq:6} \\
\Delta_{n}(Z,N) & = & \frac{(-1)^{N}}{2}( B(Z,N+1)- \nonumber \\
   & {} & 2B(Z,N)+B(Z,N-1)), \label{eq:7}
\end{eqnarray}
the $Q_{\alpha}$ value is expressed as
\begin{eqnarray}
Q_{\alpha}(Z,N) & = & Q_{pn} + Q_{pair} + Q_{S} + B_{\alpha}, \label{eq:8} \\
Q_{pn} & = & 4\delta V^{(2)}(Z,N),                            \label{eq:9} \\
Q_{pair} & = & 2( (-1)^{N}\Delta_{n}(Z,N-1) \nonumber \\
 & + & (-1)^{Z}\Delta_{p}(Z-1,N) ), \hspace{0.5cm}            \label{eq:10} \\
Q_{S} & = & 2\left( S_{n}(Z,N) + S_{p}(Z,N) \right).          \label{eq:11}
\end{eqnarray}
Since $\delta V^{(2)}$ represents the {\it p-n} correlations \cite{Kaneko1,Kaneko2}, 
the {\it p-n} component $Q_{pn}$
 corresponds to the {\it p-n} correlation energy of each 
  $\alpha$-{\it particle}.
 Here, note that the number of {\it p-n} bonds
 in an $\alpha$-{\it particle} 
 is four as illustrated in Fig. \ref{fig4}. 
The {\it p-p} and {\it n-n} pairing component $Q_{pair}$ is given by the 
proton and neutron odd-even mass differences. 

\begin{figure}[t]
\includegraphics[width=8cm,height=8cm]{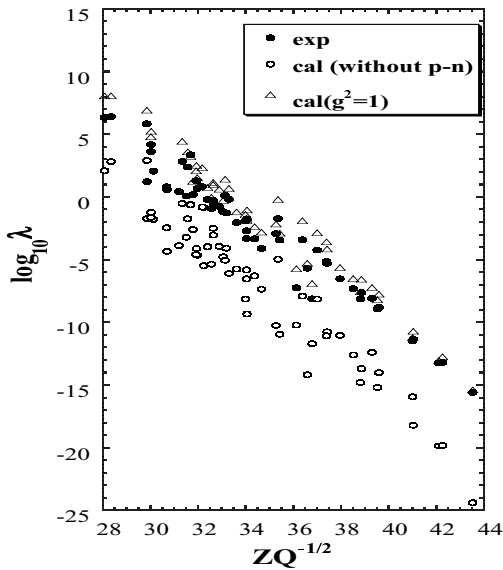}
  \caption{The ${\rm log}_{10}\lambda_{exp}$ 
              in nuclei with $Z=84\sim 100, N=110\sim 154$
              as a function of $ZQ^{-1/2}$. 
              The solid circles represent the experimental values of 
              ${\rm log}_{10}\lambda_{exp}$. The 
              open circles denote the values neglecting the {\it p-n} 
              correlation energies $4\delta V^{(2)}$.}
  \label{fig5}
\end{figure}

The upward discontinuity of $Q_{\alpha}$ value at $N=128$ is 
highest for $^{210}$Pd, $^{211}$Bi and $^{212}$Po,
 and decreases monotonously both in lighter and 
heavier elements when $|Z-82|$ increases. 
Similar behavior is observed in  systematics of separation energy, namely 
this increase comes mainly from the neutron separation energy $S_{n}$. 
The magic character of $Q_{\alpha}$ seems to be strongest at $Z=82$, $N=128$,
 though it occurs near other doubly magic or submagic nuclei. 
The special increase of the $Q_{\alpha}$ value at $N=128$ is attributed, 
in the first place, to the single-particle energy gaps in the magic nuclei.
  However, a similar systematics is also observed in the double difference
 of binding energies $\delta V^{(2)}(Z,N)$ in Figs. \ref{fig1} and \ref{fig2}.
 Since $Q_{pn}$ is proportional to $\delta V^{(2)}(Z,N)$, 
 the $p$-$n$ component $Q_{pn}$ must contribute to the $\alpha$ decay.
 In fact, if we remove $Q_{pn}$ from the experimental $Q_{\alpha}$ value 
 and assume $g_{L}(r_{c})=1.0$, 
 the common logarithm of the decay constant ${\rm log}_{10}\lambda$ 
 in the Wentzel-Kramers-Brillouin (WKB) approximation 
 is largely reduced as shown in Fig. \ref{fig5}.
 Figure \ref{fig5} shows the significant influence of $Q_{pn}$
  on the $\alpha$ decay. 
 Thus the {\it p-n} correlation energy $Q_{pn}$ increases the 
 $\alpha$-decay width, though it is smaller than the separation energy
 and the odd-even mass difference. 
 The previous analysis using the extended $P+QQ$ force model tells us 
 that $\delta V^{(2)}$ mainly corresponds to the $J$-independent $T=0$ {\it p-n} 
 interactions \cite{Kaneko1,Kaneko2}. 
 Therefore, Fig. \ref{fig5} testifies that the $\alpha$-decay transition 
 is enhanced by the {\it p-n} correlations through the $Q_{\alpha}$ value. 
 The $T=0$ {\it p-n} interaction is crucial to the $\alpha$-decay phenomenon.
 
 Although we have approximated as $g_{L}(r_{c})=1.0$ in the above consideration,
 the $\alpha$ decay width depends on the $\alpha$ formation amplitude
 $g_{L}(r_{c})$ as well as the $Q_{\alpha}$ value. 
 The $\alpha$ formation amplitude is very important for the $\alpha$
 decay from the viewpoint of nuclear structure. The almost studies of the $\alpha$ decay
 have been concentrated on this problem.
  We can get a rough estimation of the $\alpha$ decay width using $g_{L}(r_{c})=1.0$
  in the largest limit.
This assumption means a situation that an $\alpha$ particle is moving 
in potential between the daughter nucleus and the 
$\alpha$ particle. 
The values of ${\rm log}_{10}\lambda$ calculated using the experimental
$Q_{\alpha}$ values and $g_{L}(r_{c})=1.0$ are plotted also in Fig. \ref{fig5} 
where the channel radius $r_{c}$ is taken to be beyond the 
touching point of the daughter nucleus and $\alpha$ particle, that is, 
$r_{c}=1.2A^{1/3}+3.0$ fm.
The values agree quite well with the experimental ones \cite{Firestone}
in nuclei with $Z=84\sim 100, N=110\sim 154$.
In particular, the agreement is good for nuclei with large value of $ZQ^{-1/2}$.
It is notable that the effect of the $\alpha$-formation amplitude
on the $\alpha$-decay width is smaller than that of the {\it p-n} correlations 
mentioned above. 
The $\alpha$ decay is fairly well understood in terms of tunneling
 in quantum mechanics when we use the experimental $Q_{\alpha}$ values.
There are, however, still differences between the calculation and experiment. 
These discrepancies should be improved by appropriate evaluation of
 the $\alpha$-formation amplitude.
 The correlated unit, $\alpha$-{\it particle}, in the nucleus
 can be regarded as the $\alpha$ particle 
 only with some probability
 and the realistic formation amplitude is not 1.0.

\begin{figure}
\includegraphics[width=8cm,height=8cm]{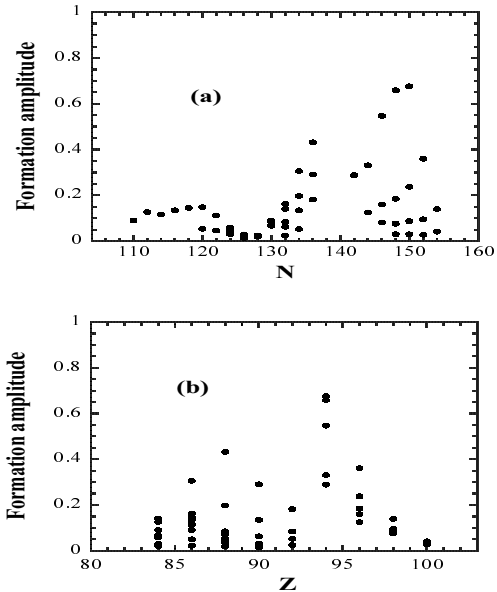}
  \caption{The formation amplitude $g^{2}$ for nuclei with proton number $Z=84\sim 100$ 
            and neutron number $N=110\sim 157$. (a) and (b) show $g^{2}$ 
              as a function of neutron number $N$ and as a function of proton 
              number $Z$, respectively. }
  \label{fig6}
\end{figure}

There are several approaches to calculation of the formation 
amplitude, the shell model, the BCS method \cite{Delion2}, the hybrid
 (shell model+$\alpha$-cluster) model \cite{Varga} etc. 
 The effect of continuum states on the $\alpha$ decay is known to be very large
 \cite{Delion2}. 
One needs therefore very large shell model basis to obtain the experimental values
of $\alpha$-formation amplitude. 
The hybrid model by Varga {\it et al.} \cite{Varga}, which treats a large shell model
basis up to the continuum states through the wavefunction of the spatially localized 
$\alpha$ cluster, explains well the experimental decay width. 
We can estimate the experimental $\alpha$-formation 
amplitude from the ratio
\begin{eqnarray}
g_{exp}^{2}(r_{c})= \frac{ \lambda_{exp} }{ \lambda_{cal} \left( g^{2}(r_{c})=1 \right) }, \label{eq:12}
\end{eqnarray}
where $\lambda_{exp}$ is the experimental $\alpha$-decay constant and 
$\lambda_{cal}( g^{2}(r_{c})=1)$ denotes the $\alpha$-decay constant
 calculated using 
  $g^{2}(r_{c})=1$ in the WKB approximation. 
 Figures \ref{fig6}(a) and \ref{fig6}(b) show the experimental $\alpha$-formation
 amplitude $g_{exp}^{2}$ as functions of $N$ and $Z$, respectively.
 A remarkable feature is that $g_{exp}^{2}$ is quit small
 when $N$ or $Z$ is a magic number, 
  and becomes larger in the middle of major shell. 
 A typical example is $g_{exp}^{2}=0.020$ in $^{212}$Po which is known
 as a spherical nucleus. 
 This value is very close to that obtained with the hybrid model \cite{Varga} and 
 the BCS approach \cite{Delion2}. 
  On the other hand, nuclei in mid-shell exhibit typical rotational spectra, 
  and are considered to be deformed nuclei. The enhancement of $g_{exp}^{2}$ 
 may be closely related to deformation. In fact, $g_{cal}^{2}$ is 
 considerably improved by introducing the deformation \cite{Delion2}, 
  while a spherical BCS method cannot explain the experiment. 
 The effect of the deformation on the $\alpha$-formation amplitude 
 seems to be remarkably large. 
    We end our discussion by commenting that the dynamical correlations due to
 the {\it p-n} interactions in addition to the static contribution to the $Q_{\alpha}$
 value are probably driving correlations of the $\alpha$-{\it particle}
  \cite{Hasegawa1}.
 
 In conclusion,
 we have shown that the nuclear correlations reveal themselves in the $\alpha$ decay
 through the $Q_{\alpha}$ value which affects the $\alpha$ penetration factor.
 We estimated the effects of the 
 {\it p-n} correlations on the $\alpha$-decay transition
 from the experimental double difference of binding energies $\delta V^{(2)}$. 
 The {\it p-n} correlations related to the $Q_{\alpha}$ value increase the 
 rate of the $\alpha$-decay transition, and plays an important role 
 particularly on the penetration process. 
 However, nuclei with $N > 128$ have large deviations from the average curve 
 $40/A$ of $\delta V^{(2)}$ which cannot be explained by the symmetry energy or 
 the $J$-independent $T=0$ {\it p-n} pairing force. This suggests that there would 
 be another interactions or correlations to describe the specific
 feature of $\delta V^{(2)}$ in this region. 
 This problem is should be studied further. 
 The ``$\alpha$-condensate" point of view suggests that the strong {\it p-n}
 correlations in $A>208$ nuclei cause the $\alpha$-like $2p-2n$
 correlations. The $\alpha$-like correlations are important
 for the penetration as well as the formation of $\alpha$-{\it particle}. 


%



\begin{references}
\bibitem{Gamov} G. Gamov, Z. Phys. {\bf 51}, 204(1928).
\bibitem{Mang} H.J. Mang, Phys. Rev. {\bf 119}, 1069(1960).
\bibitem{Arima} A. Arima and S. Yoshida, Nucl. Phys. {\bf A 219}, 475(1974).
\bibitem{Fliessback} T. Fliessback and H.J. Mang, Nucl. Phys. {\bf A 263}, 
                     75(1976). 
\bibitem{Tonozuka} I. Tonozuka and A. Arima, Nucl. Phys. {\bf A 323}, 
                  45(1979). 
\bibitem{Catara1} F. Catara, A. Insolia, E. Maglione, and A. Vitturi, Phys. 
                  Rev. {\bf C 29}, 1091(1984). 
\bibitem{Pinkston} W. T. Pinkston, Phys. Rev. {\bf C 29}, 1123(1984). 
\bibitem{Catara2} F. Catara, A. Insolia, E. Maglione, and A. Vitturi, Phys. 
                  Lett. {\bf B 149}, 41(1984).
\bibitem{Dodig1} G. Dodig-Crnkovi${\rm c}$, F. A. Janouch, R. J. Liotta, and 
                 L. J. Sibanda, Nucl. Phys. {\bf A 444}, 419(1985).
\bibitem{Dodig2} G. Dodig-Crnkovi${\rm c}$, F. A. Janouch, and R. J. Liotta,
                 Nucl. Phys. {\bf A 501}, 533(1989).
\bibitem{Delion1} D. S. Delion, A. Insolia, and R. J. Liotta, Nucl. Phys. 
                  {\bf A 549}, 407(1992). 
\bibitem{Delion2} D. S. Delion, A. Insolia, and R. J. Liotta, Phys. Rev. 
                  {\bf C 54}, 292(1996). 
\bibitem{Varga} K. Varga, R. G. Lovas, and R. J. Liotta, Nucl. Phys. 
                {\bf A 550}, 421(1992). 
\bibitem{Goodman} A. L. Goodman, Adv. Nucl. Phys. {\bf 11}, 263(1979). 
\bibitem{Zhang} J. -Y. Zhang, R. F. Casten, and D. S. Brenner, Phys. Lett. {\bf B 227}, 1(1989). 
\bibitem{Brenner} D. S. Brenner, C. Wesselborg, R. F. Casten, D. D. Warner, 
and  J. -Y. Zhang, Phys. Lett. {\bf B 243}, 1(1990).
\bibitem{Kaneko1} K. Kaneko, M. Hasegawa, Phys. Rev. {\bf C 60}, 24301(1999). 
\bibitem{Hasegawa} M. Hasegawa and K. Kaneko, Phys. Rev. {\bf C 59}, 
                   1449(1999). 
\bibitem{Kaneko2} K. Kaneko, M. Hasegawa, Prog. Theor. Phys. {\bf 106}, 1179(2001).
\bibitem{Gambhir}   Y. K. Gambhir, P. Ring, and P. Schuck, Phys. Rev. Lett. {\bf 51}, 
                    1235(1983).
\bibitem{Hasegawa1} M. Hasegawa and K. Kaneko, Phys. Rev. {\bf C 61}, 037306(2000). 
\bibitem{Dussel} G. G. Dussel, R. J. Liotta, and R. P. J. Perazzo, Nucl. Phys. {\bf A} 388, 606(1982). 
\bibitem{Roeckl} E. Roeckl, Nucl. Phys. {\bf A 400}, 113c(1983).
\bibitem{Geiger} H. Geiger and J. M. Nuttall, Phil. Mag. {\bf 22}, 613(1911). 
\bibitem{Condon} E. V. Condon and R. Gurney, Nature, {\bf 122}, 439(1928). 
\bibitem{Data} Y. A. Akovali, Nucl. Data Sheets {\bf 84}, 1(1998).
\bibitem{Delion3} D. S. Delion and A. S$\breve{a}$ndulescu, J. Phys. G {\bf 28}, 617(2002).
\bibitem{Thomas} R. G. Thomas, Prog. Theor. Phys. {\bf 12}, 253(1954). 
\bibitem{Leander} G. A. Leander, Phys. Rev. Lett. {\bf 52}, 311(1984). 
\bibitem{Gambhir1} Y. K. Gambhir, P. Ring, and P. Schuck, Phys. Rev. Lett. {\bf 52}, 312(1984). 
\bibitem{Firestone}
 \textit{Table of Isotopes}, 8th ed. by R.B.~Firestone and
 V. S. Shirley (Wiley-Interscience New York, 1996).

\end{references}
\end{document}